\documentclass[sigconf]{acmart}

\usepackage{booktabs} % For formal tables

\setcopyright{none}
\usepackage{color}

\newcommand{\sitename}{eBay} % COMMENT THIS FOR DOUBLE BLIND REVIEW
\newcommand{\eat}[1]{}

\usepackage[utf8]{inputenc}
\usepackage[english]{babel}
\usepackage{amsmath}
\usepackage{amsfonts}
\usepackage{amsthm}
\usepackage{dsfont}
\usepackage{makecell}
\usepackage{array}
\newcolumntype{?}{!{\vrule width 1pt}}

\newcommand{\freqcp}{n_{CP}}
\newcommand{\freqd}{n_{D}}
\newcommand{\probcp}{P_{CP}}
\newcommand{\probd}{P_{D}^{\sqrt{}}}

\renewcommand\footnotetextcopyrightpermission[1]{} % removes footnote with conference information in first column
\hypersetup{draft}
\begin{document}
\title{Deep Item-based Collaborative Filtering for Sparse Implicit Feedback}
%\titlenote{Produces the permission block, and
%  copyright information}
%\subtitle{Extended Abstract}
%\subtitlenote{The full version of the author's guide is available as
%  \texttt{acmart.pdf} document}

%\eat{ % UNCOMMENT THIS FOR DOUBLE BLIND REVIEW
\author{Daniel A. Galron}
\affiliation{%
  \institution{eBay Inc.}
  %\streetaddress{New}
  \city{New York} 
  \state{New York} 
  \postcode{10011}
}
\email{dgalron@ebay.com}

\author{Yuri M. Brovman}
\affiliation{%
  \institution{eBay Inc.}
  %\streetaddress{New}
  \city{New York} 
  \state{New York} 
  \postcode{10011}
}
\email{ybrovman@ebay.com}

\author{Jin Chung}
\affiliation{%
  \institution{eBay Inc.}
  %\streetaddress{New}
  \city{New York} 
  \state{New York} 
  \postcode{10011}
}
\email{jinchung@ebay.com}

\author{Michal Wieja}
\affiliation{
  \institution{eBay Inc.}
  \city{New York}
  \state{New York}
  \postcode{10011}
}
\email{michal.wieja@gmail.com}

\author{Paul Wang}
\affiliation{%
  \institution{eBay Inc.}
  %\streetaddress{New}
  \city{New York} 
  \state{New York} 
  \postcode{10011}
}
\email{pauwang@ebay.com}

%} % UNCOMMENT THIS FOR DOUBLE BLIND REVIEW
%\numberofauthors{4} % UNCOMMENT THIS FOR DOUBLE BLIND REVIEW
% \author{[Authors omitted for double blind review]} % UNCOMMENT THIS FOR DOUBLE BLIND REVIEW
%\renewcommand{\shortauthors}{D. Galron et al.}

\begin{abstract}
Recommender systems are ubiquitous in the domain of e-commerce, used to improve the
user experience and to market inventory, thereby increasing revenue for the site.  Techniques such
as item-based collaborative filtering are used to model users' behavioral interactions with items
and make recommendations from items that have similar behavioral patterns.  However, there are
challenges when applying these techniques on extremely sparse and volatile datasets.  On some
e-commerce sites, such as \sitename, the volatile inventory and minimal structured information about
items make it very difficult to aggregate user interactions with an item. In this work, we describe
a novel deep learning-based method to address the challenges.
We propose an objective function that optimizes a similarity measure
between binary implicit feedback vectors between two items.  We demonstrate formally and empirically
that a model trained to optimize this function estimates the log of the cosine similarity between
the feedback vectors.  We also propose a neural network architecture optimized on this objective.
We present the results of experiments
comparing the output of the neural network with traditional item-based collaborative filtering
models on an implicit-feedback dataset, as well as results of experiments comparing different neural
network architectures on user purchase behavior on \sitename.  Finally, we discuss the results of an
A/B test that show marked improvement of the proposed technique over \sitename's existing
collaborative filtering recommender system.
\end{abstract}




\maketitle

\section{Introduction}\label{sec:intro}
Recommender systems have become an integral part of many online services, improving their users' experience by exposing them to content they may find relevant but of which they are not yet aware.  On e-commerce sites, recommender systems are utilized to help buyers quickly filter through potentially enormous inventories to allow them to easily find items they would like to purchase.  These systems do so by constructing models of user behavior that are used to predict whether a user is likely to engage with an item.

Two widely used techniques for modeling user affinity for items are content-based filtering and collaborative filtering.  In content-based filtering, the items to be recommended are represented by characteristic features of the entities.  These may include textual features (e.g. structured tags describing the items) \cite{reference:rsh:LopsGS11}, learned image representations \cite{McAuley:2015:IRS:2766462.2767755,Jing:2015:VSP:2783258.2788621}, and audio signals \cite{ref:spotify}.  Users' preference for characteristic features are then modeled as a \emph{user profile}, and at the recommendation stage, items whose content matches a user's profile are then surfaced to the user. 

In collaborative filtering (CF) methods, recommendations are made by aggregating many users' preferences for items, and then using the aggregated preferences to make predictions for individual users.  A common method for doing so, popularized by \cite{ref:linden2003amazon,ref:deshpande2004}, is item-based collaborative filtering.  In item-based CF, items are represented by vectors indicating each user's preference for that item.  These preferences might be explicit (e.g., a rating of 1-5 indicating the degree of a user's preference), or implicit (e.g., a binary vector indicating whether a user has purchased an item or not).  A similarity measure, such as cosine similarity or the Pearson correlation coefficient is used to measure the similarity of user preferences between the items.  To provide the \emph{top-N recommendations} for a user, the similarity measure of user preferences between items is used to find the \emph{N} most similar items to those for which a given user has already expressed a preference.  
This results in an $N$-best list of items that a user has not yet engaged with, that have similar aggregate user preferences to those items that the user has engaged with.

While having shown superior performance to content-based methods, there can be challenges in applying collaborative filtering methods.  First, they have difficulty when modeling sparse user preference data.  An item must have a sufficient number of user interactions before it can be meaningfully compared with other items.  For item-based CF in particular, the general assumption is that there are more users than items.  The problem of sparsity can lead to the cold-start problem, where items new to the system cannot be used for recommendation.  Model-based approaches to collaborative filtering (e.g. \cite{ref:korenEtAl2009,ref:salakhutdinovMinh2011,ref:huEtAl2008}) oftentimes help recommendation quality on sparse datasets by learning latent representations of items only on the existing rating, while ignoring the missing data.  However, these systems still suffer from the cold-start problem, and while they are better at handling sparse data than item-based CF, there still need to be a sufficient number of ratings for an item.

% GALRON: Add Hybrid recommender systems here

While effective for e-commerce sites with fixed inventory and large user bases, the challenges suffered by item-based and model-based CF approaches can limit their applications in several recommendation settings.  At some e-commerce sites, like \sitename, the inventory consists of approximately one billion live listings at any given time, with 160 million active users.  Much of the inventory is volatile -- most items are live on site for a few weeks before they are purchased.  Furthermore, over half of the live listings are single quantity, that is, they can be purchased by at most one buyer, making the implicit user preference data (i.e., clicks and purchases) extremely sparse.  With millions of items listed daily, the cold start problem affects a substantial portion of the inventory.

Although most listings are single quantity, oftentimes they have content descriptions and metadata that may allow us to map listings to static entities, where the listings that map to a single static entity are deemed to be equivalent.  If, for example, a listing's seller provides a UPC or other manufacturer's identifier, we can map the item to a unique product ID associated with the identifier.  Item-based CF can then be applied on the product IDs to find products with similar user purchase patterns.  While this technique is effective for the top 20\% most popular purchased products on \sitename, the implicit user preferences on the product ID level are still very sparse.  Due to the heavy-tailed nature of the distribution of purchases, recommendation quality (as evaluated by user judgment and operational metrics) on the remaining 80\% of user-purchased products is low.  For the remaining listings, we can map them to other types of static entities and perform CF on them.  For example, we can use sets of \emph{aspects}, or key-value properties of items, to construct static entities, and assign listings to the entity corresponding to the set of aspects for that listing.  While the static entities constructed in this manner are very granular, and one can have a high degree of confidence that the listings that fall within an entity are equivalent items, the resulting behavioral signal is still too sparse.  The majority of entities constructed in this manner still only have one purchase for each static entity.  One can reduce the granularity of the static entities, by selecting subsets of aspects or single aspects for constructing entities.  While this approach causes the purchase signal to be more dense, it comes at the expense of the specificity of the static entities; not all items within a static entity are truly equivalent.  Another approach is to cluster listings based on the titles, and use CF on the clusters.  However, applying clustering to \sitename's data is a challenge, both because of the scale and the need to tune the granularity of the clusters appropriately for the collaborative filtering objective.

The primary challenge in all these approaches is that the mapping of listings to static entities must be constructed to optimize the recommendation quality.  The technique of constructing static entities first and then performing CF on the entities makes it difficult because the construction phase is removed from the measure of recommendation quality.  Constructing static entities can be very time consuming, and it is often not feasible to try all combinations of aspects or item clustering hyper-parameters and evaluate recommendation quality for each entity.  Instead, in this work we propose an end-to-end method that jointly learns representations of listings and uses them to predict a similarity measure of user preferences on items.  The representations, which are based on embeddings of the content features of the listings, can be thought of as mapping items to implicit entities, and estimating the user preference similarity between items based on the representations.  Formally, this model takes the form of a smooth function $h: (\phi(s), \phi(r)) \rightarrow \text{sim}(s, r)$ that embeds a \emph{seed} item $s$ (e.g. a recently purchased item) and a recommendation \emph{candidate} $r$ into a latent continuous-valued vector space based on their content features $\phi$.  The function $h$ estimates a proxy for the cosine similarity between the implicit user preference vectors between representations $\phi(s)$ and $\phi(r)$. 

The primary contributions of this work are:
\begin{itemize}
	\item An objective function whose optimal value is a monotonic transformation of the cosine similarity of implicit feedback vectors of two items
	\item A neural network architecture that embeds items into a real-valued vector space based on their content features, and optimizes the latent vectors on the cosine similarity objective
	\item A quantitative evaluation of the effectiveness of the objective function and the neural network architecture
\end{itemize}

The rest of the paper is organized as follows: in Section \ref{sec:objective} we describe our objective function, demonstrate that it optimizes the cosine similarity of implicit feedback vectors, and present a Monte Carlo method for optimizing the function; Section \ref{sec:architecture} presents our neural network architecture and some variants that allow us to learn a latent representation of items; in Section \ref{sec:related} we present some related work on applying neural networks for recommendation system and information retrieval tasks; in Section \ref{sec:experiments} we present empirical evaluations of our objective function and neural network architecture, including empirical evidence that the objective function in Section \ref{sec:objective} converges to the cosine similarity; Section \ref{sec:qualitative} describes some qualitative properties of the learned latent item representations; and in Section \ref{sec:conclusions} we describe some open questions and on-going work.

% 
% - What are the challenges of applying item-based CF to volatile inventory?
	% - Learning mappings of items to equivalent sets
	% - Collaborative filtering on feature-based static entities
	% - Solve the item cold start problem
% - Solution
	% - New objective function
	% - Combining information in titles, aspects, categories
% - In this paper...
	% - Section 2: Related works
	% - Section 3: Objective function
	% - Section 4: Neural network architectures
	% - Section 5: Experiments
	% - Section 6: Studying item embeddings
	% - Section 7: Conclusion

\section{Objective Function}\label{sec:objective}
We consider a common technique for item-based collaborative filtering \cite{ref:deshpande2004,ref:linden2003amazon}.  First, a model is built that captures the similarity of implicit feedback between the entities to be recommended.  Then, the model is applied to generate the top-N recommendations for an active user or for a seed item.  One common method of capturing the similarity is by computing the pair-wise cosine similarity of implicit feedback vectors.
Consider two items $s$ and $r$.  Let $\vec{s}$ and $\vec{r}$ be vectors of dimensionality $|U|$, where $U$ is the set of users in the system. $\vec{s}$ and $\vec{r}$ are vectors of user feedback.  A common way of measuring behavioral similarity is by computing the cosine similarity between these two vectors:
\begin{equation*}
\text{sim}(\vec{s},\vec{r}) = \cos(\vec{s}, \vec{r}) =
\dfrac
{\sum_{i=1}^{|U|} s_i \cdot r_i}
{\sqrt{\sum_{i=1}^{|U|}s_i^2} \cdot \sqrt{\sum_{i=1}^{|U|}r_i^2}}
\end{equation*}
In the case where the user feedback is implicit, and the vectors $\vec{s}$ and $\vec{r}$ are represented as bit vectors, the cosine similarity is equivalent to the \emph{Ochiai coefficient}:
\begin{equation}\label{eq:cosine}
\cos(\vec{s}, \vec{r}) =
\dfrac
{\sum_i \mathds{1}_{s_i \wedge r_i}}
{\sqrt{\sum_i s_i} \cdot \sqrt{\sum_i r_i}}
\end{equation}
Our goal is to learn a function that can estimate the cosine similarity between the implicit feedback vectors of items, based on content properties of those items. 

First, we define some of our notation.  Let $\mathcal{I}$ denote the set of all items, $t \in \mathcal{I}$.  Our training set consists of two sets: a set of item pair co-purchase transactions $(s_j, r_j) \in CP$, and a set of purchased items $t_j \in D$. The set $CP$ represents the set of transaction pairs $(s_j, r_j)$, where each pair represents an event where the same user $j$ purchased both items $s \in \mathcal{I}, r \in \mathcal{I}$. Similarly, let the set $D$ represent the set of transactions $t_j \in D$, which is the event that a user $j$ purchased an item $t \in \mathcal{I}$. Without loss of generality, assume each user will purchase an item $t \in \mathcal{I}$, or a pair of items $(s, r) \in \mathcal{I} \times \mathcal{I}$ no more than once.  We define the number of times a pair of items $(s, r) \in \mathcal{I} \times \mathcal{I}$ has been purchased by the same user as
\begin{equation*}
\freqcp(s, r) = \sum_{(x,y) \in CP}(\mathds{1}_{x=s \wedge y=r})
\end{equation*}
The total number of co-purchased pairs is then given by $|CP| = \sum_{s,r}(\freqcp(s,r))$.  Similarly, the number of times an item $t \in \mathcal{I}$ has been purchased is given by 
\begin{equation*}
\freqd(t) = \sum_{x \in D}(\mathds{1}_{x=t})
\end{equation*}
and the total number of purchases $|D|$ is given by $|D| = \sum_{s}(\freqd(s))$.

Let $h(s, r)$ be a be a parameterized function (i.e. a model) that estimates the cosine similarity of implicit feedback of items $s, r \in \mathcal{I}$.  Given a training set of co-purchased item pairs and purchased items $\mathcal{T} = (CP, D)$, we define the following cost function over the training set:
\begin{equation}\label{eq:globalloss}
\begin{split}
\ell = \sum_{(s, r) \in CP} & \freqcp(s, r) \log  \sigma(h(s, r)) \\
& + \sum_{s' \in \mathcal{I}} \sum_{r' \in \mathcal{I}} \Big[ \sqrt{\freqd(s')} \sqrt{\freqd(r')} \log \big[1 - \sigma(h(s', r')) \big] \Big]
\end{split}{}
\end{equation}
where $\sigma$ is the sigmoid function
\begin{equation*}
\sigma(x) = \dfrac{1}{1 + e^{-x}}
\end{equation*}
Note that Equation \ref{eq:globalloss} is the well-known binary cross-entropy loss applied to $\mathcal{T}$ with a particular distribution of positive and negative labels.  This loss function resembles negative sampling approaches used in word2vec and in other approaches to recommendation, such as \cite{ref:mikolovEtAl2013b,ref:covingtonEtAl2016,Xu:2016:TPR:2983323.2983874}.

We demonstrate that the value of $h(s, r)$ that minimizes this cost function for a given pair $(s, r)$ is the log of the cosine similarity expressed in Equation \ref{eq:cosine}.  Assuming the capacity of the model $h$ is large enough to allow exact prediction on $(s, r)$ without deviation from the optimum, each $h(s, r)$ can assume a value independently of other $(s, r)$ pairs.
Decomposing the loss and calculating it on a single pair of items $(s, r)$, we get the following function for a pair:
\begin{equation}\label{eq:pairloss}
\ell(s, r) = \freqcp(s, r) \log \sigma(h(s, r)) + \sqrt{\freqd(s)}\sqrt{\freqd(r)} \log \sigma(-h(s, r))
\end{equation}
Note that for convenience, here we use the fact that $1 - \sigma(x) = \sigma(-x)$.  To find the value of $h(s,r)$ that optimizes Equation \ref{eq:pairloss}, we take the partial derivative of $\ell(s, r)$ with respect to $h(s,r)$:
\begin{equation*}
\frac{\partial \ell(s,r)}{\partial h(s,r)} =  \freqcp(s, r) \sigma(-h(s,r)) - \sqrt{\freqd(s)}\sqrt{\freqd(r)}\sigma(h(s,r))
\end{equation*}
Setting this equal to 0 and solving for $h(s, r)$, we get
\begin{equation*}
\begin{aligned}
& 0 = \freqcp(s, r) \sigma(-h(s,r)) - \sqrt{\freqd(s)}\sqrt{\freqd(r)}\sigma(h(s,r)) \\
\Rightarrow & 0 = \freqcp(s, r) \frac{1}{1 + e^{h(s,r)}} - \sqrt{\freqd(s)}\sqrt{\freqd(r)}\frac{1}{1 + e^{-h(s,r)}} \\
\Rightarrow & \freqcp(s, r)\frac{1}{1 + e^{h(s,r)}} = \sqrt{\freqd(s)}\sqrt{\freqd(r)}\frac{1}{1 + e^{-h(s,r)}} \\
\Rightarrow & \frac{\freqcp(s,r)}{\sqrt{\freqd(s)}\sqrt{\freqd(r)}} = \frac{1+e^{h(s,r)}}{1+e^{-h(s,r)}}
\end{aligned}
\end{equation*}
Using the fact that $\frac{1+e^x}{1+e^{-x}} = e^x$, we find that the value of $h(s,r)$ that minimizes Equation \ref{eq:pairloss} is
\begin{equation}\label{eq:optimum}
h(s,r) = \log \Big[\frac{\freqcp(s, r)}{\sqrt{\freqd(s)}\sqrt{\freqd(r)}}\Big]
\end{equation}
This is the log of Equation \ref{eq:cosine}.  We will validate this result empirically in Section \ref{ssec:experiments:synthetic}.

Oftentimes, the cardinality of the set $\mathcal{I} \times \mathcal{I}$ is too large to be able to explicitly enumerate.  As an alternative to optimizing Equation \ref{eq:globalloss}, we can optimize a Monte Carlo estimate.  We can define the normalization factor $\mathcal{Z} \equiv \sum_{t\in\mathcal{I}} \sqrt{\freqd(t)}$.  Let $k_{CP}$ be the number of co-purchased item pairs we sample according to the distribution $\probcp(s,r) = \frac{\freqcp(s,r)}{|CP|}$.  Let $k_s$ be the number of seed items we sample as negative examples according to the distribution $\probd(s) = \frac{\sqrt{\freqd(s)}}{\mathcal{Z}}$, and let $k_r$ be the number of candidate items we sample as negative examples for each negative seed item according to the distribution $\probd(r)$. Note that we draw $s$ and $r$ \textbf{independently} from the distribution $\probd$.  We can then define the Monte Carlo estimate of the cost function in Equation \ref{eq:globalloss} as
\begin{equation}\label{eq:mcloss}
\begin{split}
\ell_{MC} = & \mathop{\mathbb{E}}_{(s, r) \sim \probcp}[k_{CP} \cdot \log \sigma(h(s, r))] \\
&+ \mathop{\mathbb{E}}_{s' \sim \probd} \mathop{\mathbb{E}}_{r' \sim \probd}[k_s \cdot k_r \cdot \log \sigma(-h(s', r'))]
\end{split}
\end{equation}
We explicitly express the expectations as follows:
\begin{equation*}
\begin{split}
\ell_{MC} = & \sum_{(s,r) \in CP} \Big[\frac{\freqcp(s,r)}{|CP|} \cdot k_{CP} \cdot \log \sigma(h(s,r))\Big] \\
& + \sum_{s' \in \mathcal{I}} \sum_{r' \in \mathcal{I}} \Big[\frac{\sqrt{\freqd(s')}}{\mathcal{Z}} \cdot \frac{\sqrt{\freqd(r')}}{\mathcal{Z}} \cdot k_s \cdot k_r \cdot \log \sigma(-h(s', r'))\Big]
\end{split}
\end{equation*}
Then, for a specific pair of items $(s, r)$:
\begin{equation}\label{eq:mclosspair}
\begin{split}
\ell_{MC}(s, r) = & \freqcp(s, r) \cdot \frac{k_{CP}}{|CP|} \cdot \log \sigma(h(s, r)) \\
& + \sqrt{\freqd(s)} \cdot \sqrt{\freqd(r)} \cdot \frac{k_s}{\mathcal{Z}} \cdot \frac{k_r}{\mathcal{Z}} \cdot \log \sigma(-h(s, r))
\end{split}
\end{equation}
Using the same methods we used to derive Equation \ref{eq:optimum} above, we can compute the derivative of $\ell_{MC}(s,r)$ with respect to $h(s,r)$ and solve for $h(s,r)$:
\begin{equation*}
\begin{split}
\frac{\partial \ell_{MC}(s, r)}{\partial h(s,r)} = & \freqcp(s, r) \cdot \frac{k_{CP}}{|CP|} \cdot \sigma(-h(s,r)) \\
& - \sqrt{\freqd(s)} \cdot \sqrt{\freqd(r)} \cdot \frac{k_s \cdot k_r}{\mathcal{Z}^2} \cdot \sigma(h(s,r))
\end{split}
\end{equation*}
Solving this for $h(s, r)$ gives us
\begin{equation}\label{eq:mc_optimum}
h(s,r) = \log \Big[\frac{\freqcp(s, r)}{\sqrt{\freqd(s)}\sqrt{\freqd(r)}}\Big] + \log \Big[\frac{k_{CP}}{k_s\cdot k_r}\Big] + \log\Big[\frac{\mathcal{Z}^2}{|CP|}\Big]
\end{equation}

Equation \ref{eq:mc_optimum} indicates that the output of the model $h(s,r)$ that optimizes the cost function in Equation \ref{eq:mcloss} is the cosine similarity shifted by a constant proportional to the ratio of the sampling mixture of positive and negative examples, and the ratio of the number of co-purchases in the training set and the number of purchases.  In Section \ref{ssec:experiments:synthetic}, we empirically demonstrate the effects of different sampling ratios $k_{CP} / k_s \cdot k_r$.

\section{Model Architectures}\label{sec:architecture}
In this section, we describe two neural network architectures that we experiment with.  These architectures define the function $h(s,r)$ whose cost function in Section \ref{sec:objective} we optimize.  The networks take two items as input, encoded as sparse feature vectors.  The neural network architecture can be decomposed into three sub-networks - one that computes a representation of item $s$, one that computes a representation of item $r$, and one that makes a prediction based on the embeddings of items $s$ and $r$.  We first describe the architecture of the sub-networks responsible for constructing embeddings for items $s$ and $r$, and then discuss how we combine the two embeddings to make a prediction.

\subsection{Item Embedding Network}\label{ssec:embedding}

For each item $t \in \mathcal{I}$, we assume we have $k$ sets of tokens associated with item $t$.  One set of tokens might be a title of an item $t$; another set might consist of aspect tokens; another might consist of category information; etc.  We make no a-priori assumptions about the nature of the features in each set, other than a constraint that they are text tokens.  We denote each set as $t_k$, and a token index $i$ in the set $t_k$ as $t_{ki}$.  Note that each item $s, r \in \mathcal{I}$ that serve as input to $h$ might have separate sets of token features $s_k$, $r_k$.

\begin{figure}
\centering
\includegraphics[width=\linewidth]{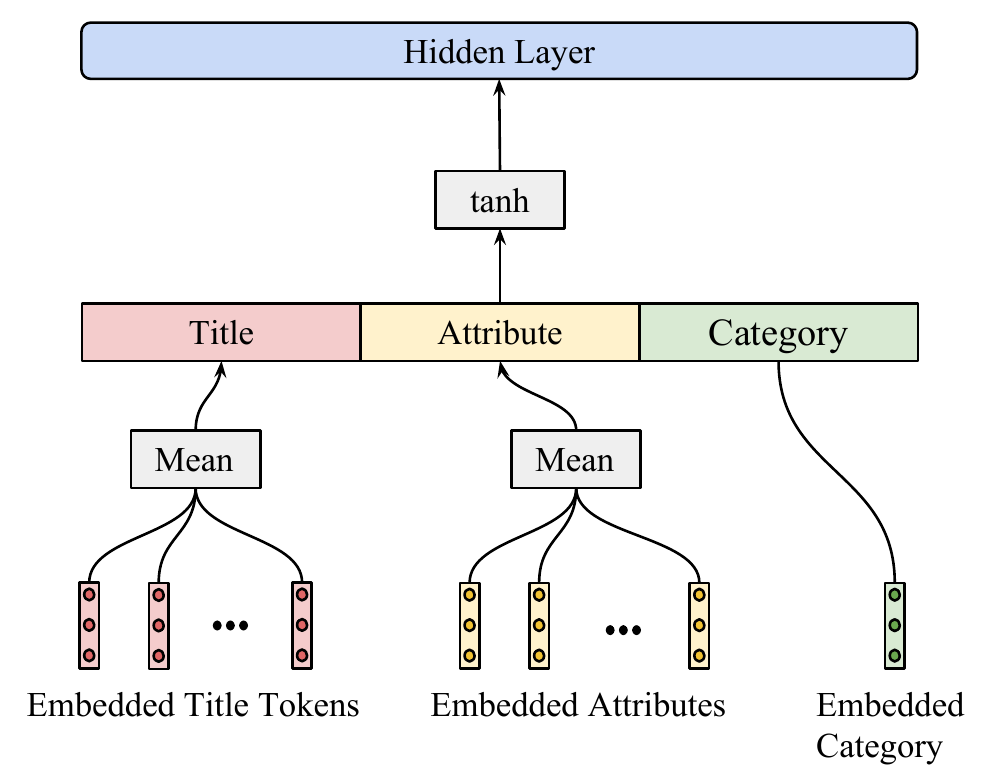}
\caption{DCF-Mean item embedder}
\label{fig:dcfmean}
\vspace{-5mm}
\end{figure}

Inspired by work on word embeddings \cite{ref:mikolovEtAl2013a,ref:mikolovEtAl2013b}, we first embed each token into a lookup table, where we define a separate lookup table for each set of tokens $t_k$.  A pooling layer then aggregates together the embedding vectors for each token $t_{ki}$ present in an item $t$ for each set $t_k$, yielding an embedding for each $t_k$.  The aggregated vectors are then concatenated together for each $t$, and fed through a non-linear activation function.  We then use a feed-forward layer to combine information between the different pooled vectors $t_k$, and finally pass the resulting vector through a non-linear activation function.  A schematic of this approach, using a mean-pooling mechanism to pool together the token embeddings, can be seen in Figure \ref{fig:dcfmean}.  Note that in this work, we use $\tanh$ as the activation function.  We experimented with ReLU activations as well, but found that they gave similar performance to $\tanh$.

The choice of vector pooling mechanism can depend on the nature of the set $t_k$.  We experiment with two methods.  When we have non-sequential text features, such as tags or aspects, we apply an element-wise mean pooling layer for the vectors associated with each token $t_{ki}$.  When we have sequential text features, such as titles, where the order of tokens may be relevant, we experiment with a few options -- treating each sequence as a bag of words and applying element-wise mean pooling; or applying a recurrent neural network (RNN) to learn a representation of the title.  We refer to the mean-pooling architecture as \textbf{DCF-Mean} (where DCF is an abbreviation for deep collaborative filtering), and the RNN architecture as \textbf{DCF-RNN}.  When applying a recurrent neural network to aggregate the tokens in a text sequence, we treat the RNN as an \emph{encoder} \cite{ref:choEtAl2014} and use the hidden state of the last time step as the embedding of the token sequence.  The end-to-end neural network shown in Figure \ref{fig:endtoend} shows the DCF-RNN variant of the item embedders.  In Section \ref{ssec:experiments:ebay}, we present results on both approaches.

\subsection{Prediction Network}\label{ssec:prediction}
\begin{figure*}
\centering
\includegraphics[width=\textwidth,height=4in]{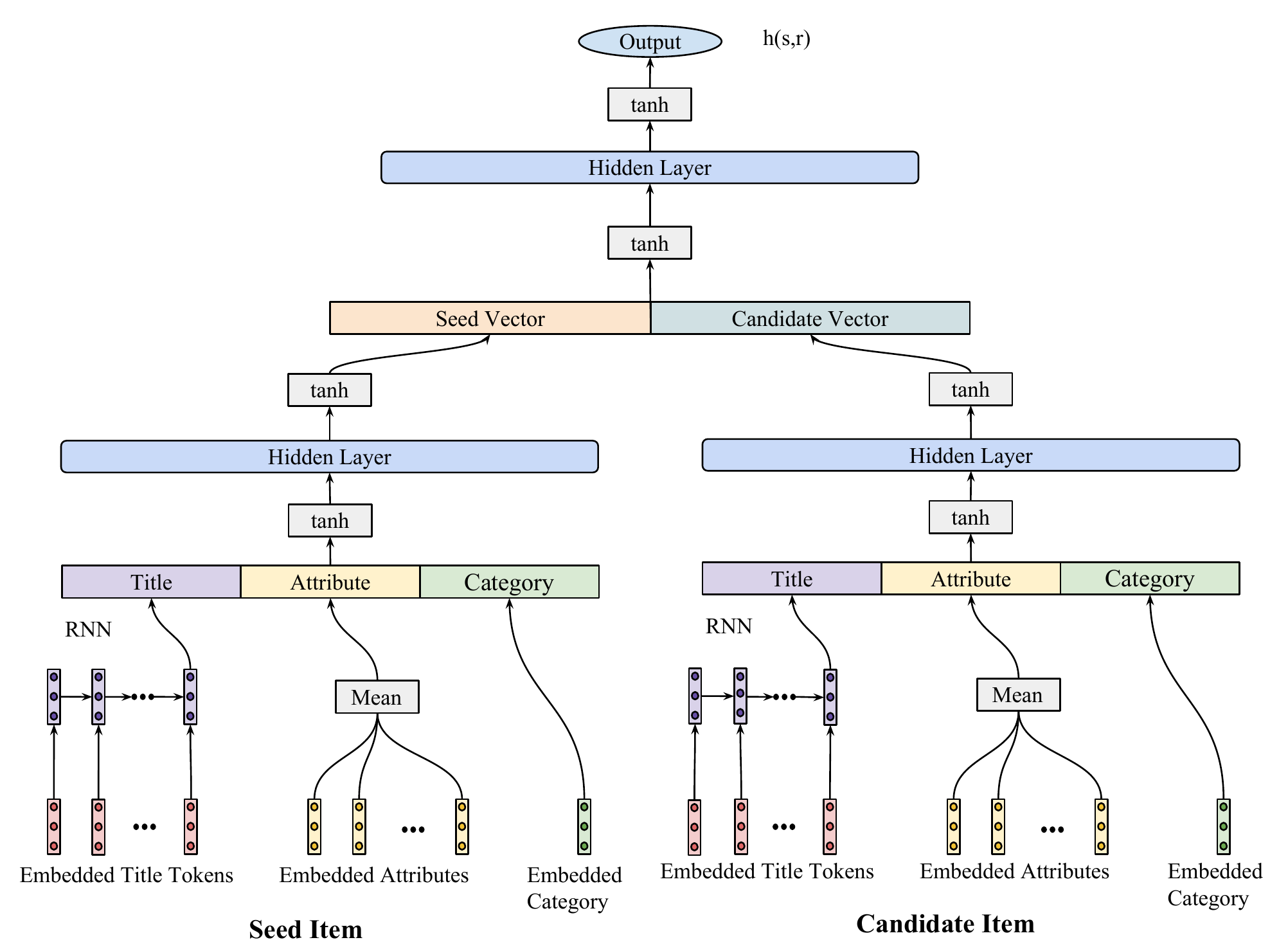}
\caption{End-to-end neural network architecture modeling $h(s,r)$}\label{fig:endtoend}
\vspace{-5mm}
\end{figure*}

The component of the neural network that takes the embeddings of the two items and outputs an estimate of a similarity measure is shown in Figure \ref{fig:endtoend}.  This part of the network concatenates the latent vector representations of items $s$ and $r$, applies a fully-connected hidden layer to the concatenated vectors, applies a non-linear activation function, and finally outputs a scalar value.  The dimensionality of the token embeddings, RNN hidden state, and fully-connected layers are all hyper-parameters that must be tuned.  Note that the end-to-end architecture can be designed to be symmetric or asymmetric -- one can tie the parameters of the two subnetworks that embed the items $s$ and $r$ so that they have the same values.  Similarly, one can tie the parameters of the two halves of the matrix corresponding to the hidden layer that combines the item embeddings.

\section{Related Works}\label{sec:related}

\textbf{Hybrid Recommender Systems}\ \ \ There have been many works on incorporating content into collaborative filtering approaches.  These \emph{hybrid recommender systems} (refer to \cite{ref:jannachEtAl2010,Burke:2002:HRS:586321.586352} for a survey of the subject) use additional metadata features to augment implicit and explicit feedback signals on items.  Basilico \& Hofmann \cite{Basilico:2004:UCC:1015330.1015394} propose a kernel-based method that constructs feature maps of users and items, and then use perceptron learning to optimize the kernels.  Melville et al. \cite{Melville:2002:CCF:777092.777124} propose a method that, given a sparse user-item explicit feedback matrix, first imputes the missing ratings using a content-based recommender system, and then performs collaborative filtering using this imputed matrix.  Li \& Kim \cite{Li:2003:ACC:1118935.1118938} first cluster items based on content features to create an item group matrix, and then perform collaborative filtering on the item groups, finally applying a linear combination of the results of item-based and group-based CF. Popescul et al. \cite{Popescul:2001:PMU:2074022.2074076} describe a generative probabilistic model that models a three-way co-occurrence between users, items, and item content.  This model estimates the parameters of a latent variable based on the users, that the items and content are then conditioned on.  Our approach is similar in spirit to these works.  All these works implicitly or explicitly map items or users to representations that encode information on their content, and perform collaborative filtering on these representations.  Our approach falls within this same setting, in that we learn latent representations of items based on their content features, and then condition our predictions of the distribution of co-purchases on these latent representations.  

\textbf{Deep Learning in Information Retrieval}\ \ \ In recent years, there has been a lot of attention paid to applying deep learning for different IR tasks, including recommender systems.  In the context of recommender systems, several works have used deep learning to solve a variety of problems, including retweet prediction \cite{Zhang:2016:RPA:2983323.2983809}, tag-aware recommendation \cite{Xu:2016:TPR:2983323.2983874,Zuo:2016:TRS:2955826.2955877}, personalized recommendations \cite{Liang:2015:PRA:2806416.2806633}, and video recommendations \cite{ref:youtube}.  Deep learning systems have also been used for IR tasks including web search \cite{Huang:2013:LDS:2505515.2505665,DBLP:journals/corr/JaechKRC17,DBLP:journals/corr/Mitra0C16} and text matching \cite{Severyn:2015:LRS:2766462.2767738,Shen:2014:LSM:2661829.2661935,NIPS2014_5550,NIPS2013_5019}.  Our work is most similar to \cite{Xu:2016:TPR:2983323.2983874} and \cite{ref:youtube}.  Both works use a negative sampling scheme similar to \cite{ref:mikolovEtAl2013a} to allow their model to scale.  In our work, we also use a negative sampling scheme, but design our sampling distributions so that we converge to an estimate of the cosine similarity between implicit feedback vectors of items.

\section{Experimental Studies}\label{sec:experiments}
We designed two sets of experiments to allow us to empirically study the properties of our objective function and our neural network architectures.  In the first set of experiments, we empirically validate that optimizing the cost function in Equation \ref{eq:globalloss} converges to the log of the cosine distance between the implicit feedback vectors.  We also study the effects of the sampling ratios when optimizing the Monte Carlo estimate in Equation \ref{eq:mcloss}.  In our second set of experiments, discussed in Section \ref{ssec:experiments:ebay}, we compare the performance of DCF-Mean and DCF-RNN on \sitename's user behavioral data, and present the results of an A/B test comparing the new approach with \sitename's existing CF-based recommender system.

\subsection{Synthetic Experiments}\label{ssec:experiments:synthetic}
\textbf{Synthetic Dataset}\ \ \ To validate that a model trained to optimize the cost function in Equation \ref{eq:globalloss} optimizes the cosine similarity between implicit feedback vectors, we generated a synthetic dataset of purchases and trained a model to predict the cosine distances between the implicit vectors.  We used the following procedure to generate a dataset.  Let $N_u$ be the number of users in a system, and let $N_i$ be the number of items.  For each item $i = 1 \ldots N_i$, we draw a random variable $p_i \sim \text{Uniform}(0.2, 0.8)$.  Then, for each user $u = 1 \ldots N_u$ and for each item $i$, we draw a random variable $r_{ui} \sim \text{Bernoulli}(p=p_i)$ indicating whether $u$ has purchased item $i$.  This gives us an implicit feedback matrix of dimensionality $N_u \times N_i$.  We then construct a set of training examples from this user-item feedback matrix.  For each pair of items $i, j$, we construct a feature vector of dimensionality $N_i \cdot N_i$, where all entries are set to 0 except for the entry $k=i*N_i + j$, which is set to 1.  This has the effect of creating a feature for each pair of items $(i, j)$.  For all the experiments in this section, the synthesized dataset consisted of 100 items and 10,000 users.  There were 25,355,794 total co-purchases in the synthesized dataset, with a sparsity ratio of 25.35\%.  We experimented with several different values for the number of items, number of users and sparsity, but they made little difference in the results of the experiments described below.  The only difference was in training time.

\begin{figure}
\centering
\includegraphics[width=\linewidth]{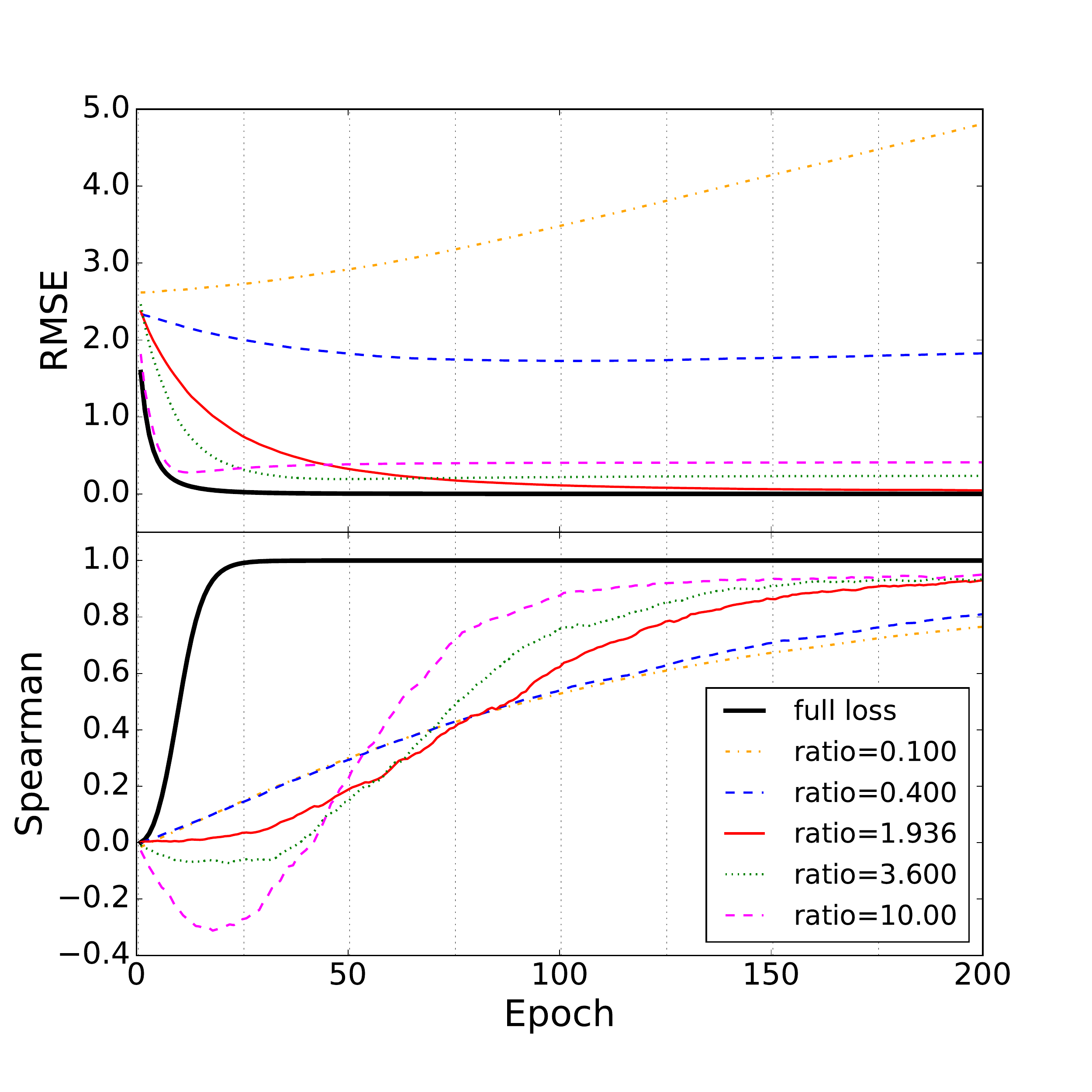}
\caption{Effects of different ratios of $\frac{k_{CP}}{k_s \cdot k_r}$ on convergence rate of Monte Carlo loss.}
\label{fig:mc_convergence}
\vspace{-5mm}
\end{figure}

\textbf{Validating Full Cost Function}\ \ \ To empirically validate the cost function presented in Equation \ref{eq:globalloss}, we trained a model on the synthetic dataset previously described.  We then compared the model's estimates of the cosine similarity between each pair of items $(i, j)$ against the true calculated cosine similarity between $i$ and $j$, using both RMSE and Spearman's rank correlation coefficient.  The training set consisted of a feature vector for each pair of items $(i, j)$, along with their co-occurrence frequencies $\freqcp(i, j)$ and the square root of their individual frequencies $\freqd(i)$.  We trained a linear model of the following form

\begin{equation*}
h(s,r) = \vec{\theta} x_{sr}
\end{equation*}
to optimize Equation \ref{eq:globalloss}.  In this case, each parameter $\theta_{k}$ in the parameter vector will correspond to the similarity of the items $(i, j)$ (since we designed our feature vectors to be indicator variables for each pair $k=i*N_i + j$).  We trained the model for 200 epochs with learning rate 0.1 using stochastic gradient descent.  As the purpose of this experiment was to empirically validate the convergence properties, we compared the model's output against the cosine similarity on the training set. At the conclusion of the training run, the RMSE between the calculation of the true cosine similarities on the training set and our model's estimates was $1.0461 \times 10^{-4}$, and the Spearman rank correlation coefficient was 0.9999, with $p < 0.01$.  Additionally, we also tracked the convergence of the model.  Those results can be found in the heavy-lined curve in Figure \ref{fig:mc_convergence}.

\textbf{Effect of Sampling Ratios}\ \ \ With very large inventories, it is often infeasible to do a full pass over all item pairs, which is a requirement for optimizing the cost function in Equation \ref{eq:globalloss}.  As an alternative, one can use a Monte Carlo technique to approximate the expectations in the cost function in Equation \ref{eq:mcloss}.  As demonstrated in Section \ref{sec:objective}, the value of the model output that minimizes Equation \ref{eq:mcloss} is the log of the cosine similarity of the implicit feedback vectors shifted by a constant:
\begin{equation*}
h(s,r) = \log \Big[\frac{\freqcp(s, r)}{\sqrt{\freqd(s)}\sqrt{\freqd(r)}}\Big] + \log \Big[\frac{k_{CP}}{k_s\cdot k_r}\Big] + \log\Big[\frac{\mathcal{Z}^2}{|CP|}\Big]
\end{equation*}
where $k_{CP}$ is the number of true co-purchases to sample, $k_s$ is the number of seed items to sample, and $k_r$ is the number of candidate items to sample.  To converge to the optimal value of Equation \ref{eq:globalloss} when using Monte Carlo techniques, this result suggests that $k_{CP}$, $k_s$, and $k_r$ ought to be selected so that the ratio of samples $\frac{k_{CP}}{k_s \cdot k_r}$ is set to $\frac{|CP|}{\mathcal{Z}^2}$, where $|CP|$ denotes the number of co-purchased item pairs, and $\mathcal{Z}$ denotes the sum of square roots of the number of purchases of each item.  Oftentimes, however, we are more interested in the ranking of similar items produced by the cosine distance rather than the values themselves. To understand the effect of the selection of $k_{CP}$, $k_s$, and $k_r$ on the convergence rate of the training algorithm, we experimented with different ratio values of $\frac{k_{CP}}{k_s \cdot k_r}$. We fixed $k_{CP} = 100000$, and tried several values of $k_s$ and $k_r$, training the model for a fixed 200 epochs, where on each epoch we sampled examples from the distributions $\probcp$ and $\probd$.  Note that for this synthetic experiment, the ratio $\frac{|CP|}{\mathcal{Z}^2} = 0.516$.  The convergence rates of different ratios are shown in Figure \ref{fig:mc_convergence}.   We ran several such experiments with a variety of sparsity levels (and therefore different ratios $\frac{|CP|}{\mathcal{Z}^2}$), and the convergence results were comparable.

As expected, setting the ratio to $\frac{|CP|}{\mathcal{Z}^2}$ yielded the lowest RMSE compared to the true cosine distances at convergence.  The result in Section \ref{sec:objective} indicates that setting the ratio to $\frac{|CP|}{\mathcal{Z}^2}$ in Equation \ref{eq:mc_optimum} gives the optimal value in Equation \ref{eq:optimum}.  Interestingly, oversampling items from $\probd$ (giving a lower ratio $\frac{k_{CP}}{k_s \cdot k_r}$) caused the Spearman rank correlation coefficient between the model output and the true cosine distance to converge more quickly, although they yielded worse RMSE values.  This suggests that if one is training a model for a fixed number of epochs, oversampling $k_s$ and $k_r$ would allow the model to reach a good value more quickly than setting the ratio to $\frac{|CP|}{\mathcal{Z}^2}$. Under-sampling $k_s$ and $k_r$, so that $\frac{k_{CP}}{k_s \cdot k_r} > \frac{|CP|}{\mathcal{Z}^2}$, caused the RMSE to diverge, and the Spearman Rank coefficient to converge significantly more slowly.

\subsection{Experiments on \sitename~Data}\label{ssec:experiments:ebay}
\textbf{Setting}\ \ \ \sitename~has several recommender systems in use for different recommendation settings.  We concern ourselves with the setting of \emph{post-purchase} recommendations, recommendations that are made after a user purchases an item (referred to as the \emph{seed} item).  The current post-purchase recommender system constructs a recall set of recommendation candidates for the seed item using item-based collaborative filtering on a few levels of static entities, and then applies a set of business rules to select the top N recommendations among the recall sets.  Note that these results are not personalized to a user based on their entire purchase history.

\textbf{Data}\ \ \ \sitename's inventory is segmented into a hierarchy of categories.  For both our off-line experiments and for the user-facing recommender system, we trained one model for each high-level category.  In this work, we present results on five of those categories: women's fashion, electronics, skin care, outdoor sporting goods, and collectibles. These categories were selected to highlight a variety of inventory.  We define the implicit feedback signals we train on as purchases -- we sample pairs of listings from $\probcp$ that were co-purchased by the same user, and we sample purchased listings from the listing distribution $\probd$.  We constructed our training sets for the five models as follows.  For each of the five categories, we collected all purchases made on \sitename~between January 1st 2015 and May 31st 2016 under those categories.  For each category, we constructed a set of co-purchased item pairs $CP$ by taking all pairs of items $(t_1, t_2)$ under the category purchased by the same user $u$, with the constraint that $t_2$ was purchased after $t_1$.  We construct the set of purchased listings $D$ by including all purchased items under the category.  Statistics about this dataset are summarized in Table \ref{tab:trainStats}.

\begin{table}[t]
\centering
\begin{tabular}{|c||c|c|c|c|}
\Xhline{3\arrayrulewidth}
\textbf{Dataset} & $|CP|$ & $\mathcal{Z}^2$ & \textbf{\# Features} \\
\Xhline{3\arrayrulewidth}
\parbox[t]{2.5cm}{\textbf{Women's Fashion}} & 34894458 & 265921110 & 29318 \\
\hline
\parbox[t]{2.5cm}{\textbf{Electronics}} & 5297580 & 42685339 & 35620 \\
\hline
\parbox[t]{2.5cm}{\textbf{Skin Care}} & 6060626 & 8986127 & 55456 \\
\hline
\parbox[t]{2.5cm}{\textbf{Outdoor Sporting}\\ \textbf{Goods}} & 2097230 & 36683099 & 51163 \\
\hline
\parbox[t]{2.5cm}{\textbf{Collectibles}} & 13904465 & 38635017 & 36169\\
\Xhline{3\arrayrulewidth}
\end{tabular}
\caption{Statistics on training datasets for \sitename}\label{tab:trainStats}
\vspace{-10mm}
\end{table}
Note that this method of constructing our training set, along with the fact that we have independent parameters for the three sub-networks in Figure \ref{fig:endtoend}, means that our models will be asymmetric -- the output of the model will not be the same if we invert the seed and candidate items.

\textbf{Evaluation Methodology}\ \ \ To construct our test set, we sampled listings from the set of purchases made in June 2016, ensuring that the sets of test and training items were completely disjoint.  From the set of June 2016 purchases, we randomly sampled 250 purchased items from each category. For each of those purchased items, we randomly sampled a subsequent purchase by the same user under the same high-level category.  We refer to these 250 pairs as the set of \emph{true co-purchases}.  When generating recommendations on \sitename, we are typically going to consider approximately 1 million recommendation candidates for each seed item.  To reflect this in our evaluation setting, we require a very large recall set of candidate items, from which we will select the top 30 recommendations.  Specifically, we randomly sampled 200,000 random items from the high-level category to serve as the recall set of candidates.  Thus, there will be 200,001 recommendation candidates for each seed item, with one of them being a true co-purchase.  The model is applied to rank this set of candidates for each seed.  Because our evaluation set contains only one true co-purchased item for every seed, we evaluate our models by measuring the rank of the true co-purchased item in the list.

We evaluate our results using two measures: mean recall at $k$ and mean reciprocal rank.  In information retrieval, recall at $k$ is the mean of the number of relevant results in the top $k$ of a ranked list divided by the total number of relevant results:
\begin{equation*}
\text{rec}_k(s) = \frac{1}{n}\sum_{i=1}^n \dfrac{\#(\text{relevant results at position} > k)}{\#(\text{total number of relevant results})} 
\end{equation*}
In our case, we have one relevant (co-purchased) item for each seed, so recall at $k$ measures the percentage of seed items for which the true co-purchase is in the top $k$ of the ranked list.  On \sitename, 30 recommendations are surfaced for a seed.  We evaluate our models in the same setting, fixing $k=30$.  To understand the performance of our model on the entire recall set and not just the top 30, we also evaluate the mean reciprocal rank of the true co-purchased item, namely:
\begin{equation*}
\text{MRR} = \frac{1}{n}\sum_{i=1}^n \frac{1}{\text{rank}(i)}
\end{equation*}
% When evaluating the performance of the model on such a large recall set, however, a few low ranks of the true co-purchase can skew the mean reciprocal rank downwards to make it difficult to understand how often the true co-purchase is found at the top of the ranked list.  As an alternative, we evaluated our model using the \emph{median} of the rank of the true co-purchased item in the ranked list.  Note that for MRR and recall at $k$ a higher value indicates better model performance, but for median rank a lower value does (since we take the rank and not its reciprocal into account).  Also note that median rank is unnormalized - however, given the fixed size of the recall set we report results are, the median rank values are comparable across categories and across experiment settings.

%% \textbf{GALRON: SAY SOMETHING ABOUT RELEVANCE}

\textbf{Training Details}\ \ \ We demonstrated in Section \ref{ssec:experiments:synthetic} that the ideal ratio $\frac{k_{CP}}{k_s \cdot k_r}$ should be set to $\frac{|CP|}{\mathcal{Z}^2}$.  However, the degree of sparsity in the training data makes it impractical to follow this guideline.  As implied by the results in Section \ref{ssec:experiments:synthetic}, and the statistics of the size of $|CP|$ and $\mathcal{Z}^2$ in Table \ref{tab:trainStats}, we would need to sample several thousand items from $\probd$ for each $s$ and $r$ for every sample from $\probcp$.  Instead of directly optimizing Equation \ref{eq:mcloss}, we note that our sampling of items $s$ from $\probd$ make no difference in the ranking order of candidates for a given seed item $s$.  Inspired by \cite{ref:mikolovEtAl2013a,ref:mikolovEtAl2013b}, we instead sample a fixed number of $r$ items for each seed $s$ from $\probd$.  The cost function we end up optimizing for these experiments is given by:

\begin{equation}\label{eq:experiment_loss}
\begin{split}
\ell =  \mathop{\mathbb{E}}_{(s, r) \sim \probcp}\Big[k_{CP} \cdot \log \sigma(h(s, r)) 
+ \mathop{\mathbb{E}}_{r' \sim \probd}[k_r \cdot \log \sigma(-h(s, r'))]\Big]
\end{split}
\end{equation}
Note the similarity between this approximation and the objective function in \cite{ref:mikolovEtAl2013a,ref:mikolovEtAl2013b}.

We compare the performance of two neural network architectures for this task -- one that uses mean pooling on the title token embeddings and the key-value property embeddings (DCF-Mean), shown in Figure \ref{fig:dcfmean}, and one that uses a vanilla recurrent neural network to embed the title, while using mean pooling to aggregate the aspect embeddings (DCF-RNN), shown in Figure \ref{fig:endtoend}.  We implemented our models and training algorithms in Theano \cite{ref:theano}, and trained the models on an NVIDIA Tesla M40 card, for a maximum of 1000 epochs.  In practice, training time ranged between three and five days.  In DCF-Mean, we fixed the token embeddings to have 200 dimensions, the hidden layer in the item embedders to have 400 dimensions, and the hidden layer in the top layer with 1200.  DCF-RNN had the same setting, with the hidden state of the RNN set to 200 dimensions as well.

We compare the performance of these two neural networks against two baselines -- a linear model baseline (i.e. a logistic regression model) trained to minimize Equation \ref{eq:experiment_loss}, and the current collaborative filtering method employed by \sitename.  The training examples for the linear model consist of binary vector representations of listings, where each item is represented by a $|V|$-dimensional vector, where $|V|$ is the size of the vocabulary for the model.  If a feature is present in the listing, its corresponding entry in the feature vector is set to 1, otherwise, it is set to 0.  As described in the introduction, the \sitename~collaborative filtering algorithm maps items to static entities and then computes the cosine similarity between implicit feedback vectors of static entities.  The highest quality static entity is a product, which we derive from the UPC or other manufacturer's identifier for an item.  The fallback static entity is an aspect, of which an item may have more than one.  For a given seed item $s$ and a set of candidate items $\{r_1, \ldots, r_k\}$, if $s$ maps onto a product ID, then we rank the subset of $\{r_1, \ldots, r_k\}$ that also map onto products by the cosine distance on purchase vectors between the products.  The ranked list of products is at the top of the $N$-best list.  For the remaining subset of $\{r_1, \ldots, r_k\}$ that is not mapped to a product, we fetch all the aspects for each item in the list and compute the cosine similarity between all aspects of the seed and all aspects of the candidate.  The score for a pair $(s, r_i)$ is then the sum of the similarities between all property pairs.  The candidates are ranked according to these scores and are appended to the list after the items mapped to products.

\textbf{Quantitative Results}\ \ \ The results of our experiments can be found in Tables \ref{tab:arch_experiments} and \ref{tab:k_experiments}.  We ran two sets of experiments.  First, we compared the performance of different model architectures trained on the same objective function on the five categories.  We compared the performance of the baseline linear model, DCF-Mean, and DCF-RNN on each of the five high-level categories.  For all of these experiments, we fixed $k_{cp} = 200000$ and $k_r=4$ for each epoch of training.  We fixed the maximum number of epochs to 1000, and used early stopping on a held out validation set to stop training when the loss on the validation set stopped decreasing.  The results on the test set are found in Table \ref{tab:arch_experiments}.  Results in bold are statistically significant at $p < 0.01$, and results in italics are statistically significant at $p < 0.05$.  Statistical significance was estimated using Student's paired t-test.
\begin{table*}[t]
\centering
\begin{tabular}{?l?c|c|c|c?c|c|c|c?}
\Xhline{3\arrayrulewidth}
Category & \multicolumn{4}{c?}{MRR} & \multicolumn{4}{c?}{Recall@30} \\
\hline
& CF Baseline & Linear & DCF-Mean & DCF-RNN & CF Baseline & Linear & DCF-Mean & DCF-RNN \\
\Xhline{3\arrayrulewidth}
\parbox[t]{2.5cm}{Women's Fashion} & 0.0063 & 0.001 & \textit{0.018} & \textit{0.024} & 0.016 & 0.016 & \textbf{0.104} & \textbf{0.100} \\
\hline
\parbox[t]{2.5cm}{Electronics} & 0.0069 & 0.0005 & 0.0055 & 0.002 & 0.020 & 0.004 & 0.040 & 0.024 \\
\hline
\parbox[t]{2.5cm}{Skin Care} & 0.0030 & 0.0001 & \textit{0.0126} & 0.0039 & 0.012 & 0.0 & \textbf{0.084} & 0.020 \\
\hline
\parbox[t]{2.5cm}{Outdoor Sporting\\ Goods} & 0.0058 & 0.0043 & 0.0057 & 0.0112 & 0.032 & 0.016 & 0.048 & 0.032 \\
\hline
\parbox[t]{2.5cm}{Collectibles} & 0.0091 & 0.0003 & 0.0112 & 0.0041 & 0.020 & 0.0 & \textbf{0.072} & 0.012 \\
\Xhline{3\arrayrulewidth}
\end{tabular}
\caption{Results comparing the performance of different model architectures. Results in \textbf{bold} indicate that the result is statistically significant over the baseline at $p < 0.01$, results in \textit{italics} are statistically significant at $p < 0.05$.}\label{tab:arch_experiments}
\vspace{-5mm}
\end{table*}

The results indicate that with $k_r=4$, DCF-Mean achieves the best ranking performance over the baseline.  DCF-Mean achieves statistically significant improvement over the CF-based baseline in three of the five categories.  Interestingly, DCF-RNN did not achieve statistically significant improvements over the baseline, despite the fact that it is able to account for token ordering in the titles.  We conjecture that our use of vanilla RNNs hampered our model's ability to learn due to the vanishing gradient problem (we also noticed that the loss plateaued earlier for the RNN models than for DCF-Mean).  In future work, we will experiment with alternative RNN formulations, such as LSTMs \cite{ref:lstm} or GRUs \cite{ref:gru}, as well as convolutional methods (e.g., \cite{Zhang:2015:CCN:2969239.2969312}).  Additionally, for two of the categories, DCF-Mean was unable to gain a statistically significant advantage over the collaborative filtering baseline.  As we show below, DCF-Mean will outperform the baseline for higher values of $k_r$.

The second set of experiments we ran compared the effect of different selections of $k_r$ on the ranking performance of DCF-Mean.  Generally speaking, the more items $r$ sampled from $\probd$, the larger the gain over the CF baseline.  Interestingly, the strongest results were in the Women's Fashion, Skin Care, and Collectibles categories.  Purchases in these categories tend to better reflect a user's taste (for example, collectors' interest in certain sports memorabilia, or buyers with particular sartorial preferences) than in the more utilitarian categories of electronics or outdoor sporting goods.  We conjecture that these categories are harder to learn because users' stylistic or thematic preferences are not as often reflected in the co-purchase data.  Additionally, it is important to note that the recall at $k$ and MRR measures only measure the position of the true co-purchased item in the ranked list.  Given the nature of the inventory on \sitename, in many settings there may be functionally equivalent items to the true co-purchased item ranked higher in the list than the true co-purchase.  It is possible that given two models with equal recall at $k$ or MRR measures on the test set, one model might rank more relevant recommendations ranked above the true co-purchase than another model, but the evaluation measures do not reflect this property.

\begin{table*}[t]
\centering
\begin{tabular}{?c?c|c|c|c|c?c|c|c|c|c?}
\Xhline{3\arrayrulewidth}
Category & \multicolumn{5}{c?}{MRR} & \multicolumn{5}{c?}{Recall@30} \\
\hline
 & CF &$k_r=1$ & $k_r=4$ & $k_r=9$ & $k_r=19$ & CF & $k_r=1$ & $k_r=4$ & $k_r=9$ & $k_r=19$ \\
\Xhline{3\arrayrulewidth}
\parbox[t]{2.5cm}{Women's Fashion} & 0.0063 & 0.016 & \textit{0.018} & \textbf{0.031} & \textbf{0.050} & 0.016 & \textbf{0.076} & \textbf{0.104} & \textbf{0.168} & \textbf{0.188} \\
\hline
\parbox[t]{2.5cm}{Electronics} & 0.0069 & 0.0018 & 0.0055 & 0.0073 & 0.0089 & 0.020 & 0.0 & 0.040 & 0.044 & \textbf{0.064} \\
\hline
\parbox[t]{2.5cm}{Skin Care} & 0.0030 & 0.0055 & \textit{0.0126} & \textbf{0.0143} & \textbf{0.0121} & 0.012 & 0.024 & \textbf{0.084} & \textbf{0.108} & \textbf{0.088} \\
\hline
\parbox[t]{2.5cm}{Outdoor Sporting\\ Goods} & 0.0058 & 0.0050 & 0.0057 & \textit{0.0114} & 0.0125 & 0.032 & 0.032 & 0.048 & 0.084 & 0.076 \\
\hline
\parbox[t]{2.5cm}{Collectibles} & 0.0091 & 0.0053 & 0.0112 & 0.0153 & 0.0153 & 0.020 & 0.024 & \textbf{0.072} & \textbf{0.096} & \textbf{0.088} \\
\Xhline{3\arrayrulewidth}
\end{tabular}
\caption{The results comparing different number of $k_r$ samples.  Results in \textbf{bold} indicate that the result is statistically significant over the baseline at $p < 0.01$, results in \textit{italics} are statistically significant at $p < 0.05$.}\label{tab:k_experiments}
\vspace{-5mm}
\end{table*}

\textbf{A/B Test Results}\ \ \ As is common on e-commerce sites, we ran an A/B test comparing our approach versus the production baseline variant to evaluate the performance of the model in an on-line setting.  The A/B test ran for two weeks on \sitename, testing the DCF-Mean model variant against the current production variant.  Table~\ref{tab:abtest} lists the results, indicating an improvement of two critical metrics: click-through rate (CTR) and purchase-through rate (PTR).  DCF-Mean demonstrated a 79.7\% increase in CTR over the baseline and a 51.8\% increase in PTR over the baseline.

\begin{table}[t]
\centering
\begin{tabular}{|c|c|c|} \hline
&CTR&PTR\\ \hline
lift&+79.7\%&+51.8\%\\ \hline
\end{tabular}
\caption{Lift in A/B test operational metrics. }
\label{tab:abtest}
\vspace{-1cm}
\end{table}

\section{Discussion}\label{sec:qualitative}
To understand why our method seems to work well in the setting of extreme sparsity and cold-start items, we can examine the representations the model learns for items.  We conjecture that the reason the model is able to perform well is due to the semantic similarities between items with similar purchase patterns.  We posit that seed items that are similar in content with each other (e.g., have similar thematic and functional semantics) are more likely to be co-purchased with candidate items that are similar in content to each other.   If true, then the item representations learned by the model ought to reflect these semantics.  While we currently do not have a quantitative measure of semantic similarity, we can examine a few examples of the representational similarity of items.

The listings are sorted by the distance between their vector embeddings (i.e., the hidden layer in Figure \ref{fig:dcfmean}) from the listing whose title is in bold.  For each item whose title is in bold, we computed the Euclidean distance between that item and every other item in our recall set.  We then sorted the list according to distance and selected the listings at the 10th, 100th, 1000th, 10000th, 20000th, 50000th, and 100000th positions.  We present the item titles in Table \ref{tab:representation} to depict their semantic similarities.

\begin{table*}
\centering
\begin{tabular}{|c|l|c|}
\hline
Position & Title & Distance \\
\hline
0 & \textbf{1992 Chicago Bulls world champions drinking glasses set of 3 NBA Michael Jordan} & 0.0 \\
10 & 1969 vintage Boston Celtics NBA championship mug stein basketball Havlicek & 0.7043 \\
100 & Mickey Mantle the Mick NY Yankees August 14 1995 news tribute & 0.9664 \\
1000 & Houston Texans die cut window decal by Rico industries & 1.1339 \\
10000 & NFL new era New England Patriots 59Fifty Team Fitted Hat size 7 3/4' & 1.2711 \\
20000 & Pittsburgh Steelers Snack Helmet by Wincraft inc. Official NFL & 1.3321 \\
50000 & 2015 Topps chrome WWE 45 Luke Harper wrestling card & 1.505 \\
100000 & 1993 Tyco vortex action figure from Double Dragon loose & 1.7590 \\
\hline
0 & \textbf{Virginia Tech Hokies cell phone hard case iPhone 4 \& 4s} & 0.0 \\
10 & Boise State Broncos cell phone hard case for Apple iPhone 4 \& 4s & 0.3031 \\
100 & Unlimited cellular phone jelly case for Apple iPhone 4 \& 4s & 0.6708 \\
1000 & Blue skull crystal black finished hard case cover for Apple iPhone 6s plus, 5, 5s & 0.7755 \\
10000 & Nike black hard phone case available for iPhone 6, 6s plus, 5, 5s fast shipping & 0.9080 \\
20000 & New 6 plus only recover wood iPhone 6 phone cover case NIB & 0.9694 \\
50000 & Magnetic gray magnetic flip wallet swede leather case for iPhone6, 6s & 1.1027 \\
100000 & The adventure of case for iPhone 4,5,6, 6s and Samsung Galaxy s4, s5, s6, s7 & 1.2506 \\
\hline
0 & \textbf{Biggs Darklighter 30th annv rebel X-Wing pilot Star Wars gold coin action figure} & 0.0 \\
10 & 2006 Star Wars Saga collection Rogue Two snowspeeder w Zev Senesca pilot figure & 0.45034447 \\
100 & 2005 Hasbro Star Wars Episode iii Revenge of the Sith Chewbacca action figure & 0.6689232 \\
1000 & Vintage Star Wars action figure 1980 lobot aid complete w weapon & 0.9111377 \\
10000 & Star Trek V Final Frontier large action figure limited & 1.0554923 \\
20000 & Harry Potter 2001 headmaster Dumbledore Sorcerers Stone action figure & 1.1852574 \\
50000 & Lightcore Hex Skylanders giants new sealed ships fast & 1.4289094 \\
100000 & 1980's original vintage New York Giants snapback baseball hat mint & 1.669067 \\
\hline
\end{tabular}
\caption{Item titles sorted by the distance between their embedding representations}\label{tab:representation}
\end{table*}

These examples suggest that the semantics of the items are represented in the distances between the vectors.  Consider the first example in Table \ref{tab:representation}: ``1992 Chicago Bulls world champions drinking glasses set of 3 NBA Michael Jordan''.  The item in the tenth position is another NBA commemorative cup, one for the Boston Celtics.  This is despite the fact that the only shared token between the two is the token ``NBA'';  the network seems to have learned to model the relationship between basketball, championships, and glasses \& mugs.  As one descends in the list, one gets baseball memorabilia, followed by football memorabilia, followed by a wrestling collectible card, followed by a collectible action figure.  One can see similar patterns for the other items.

\section{Conclusions \& Future Work}\label{sec:conclusions}
In this work, we have described a machine learning objective that converges to the cosine distance between implicit feedback vectors, allowing us to train collaborative filtering models that take into account the content similarity of the items to be recommended. We have presented two neural network architectures and demonstrated that they are more effective in the context of volatile inventory and sparse behavioral implicit feedback matrices than item-based CF, in both off-line evaluation and in an on-line A/B test.  There is substantial room for improvement however.  Finding better methods for combining the information across token vectors, such as LSTMs or convolutional neural networks, would allow us to better model the relationships between the tokens.  Additionally, we would like to extend our objective function to model the similarity in explicit feedback vectors.  Finally, we believe we can improve the model performance by exploring alternative optimization methods and item sampling strategies.

\bibliographystyle{ACM-Reference-Format}
\bibliography{bibliography} 


\begin{thebibliography}{00}

%%% the Web address.

\ifx \showCODEN    \undefined \def \showCODEN     #1{\unskip}     \fi
\ifx \showDOI      \undefined \def \showDOI       #1{#1}\fi
\ifx \showISBNx    \undefined \def \showISBNx     #1{\unskip}     \fi
\ifx \showISBNxiii \undefined \def \showISBNxiii  #1{\unskip}     \fi
\ifx \showISSN     \undefined \def \showISSN      #1{\unskip}     \fi
\ifx \showLCCN     \undefined \def \showLCCN      #1{\unskip}     \fi
\ifx \shownote     \undefined \def \shownote      #1{#1}          \fi
\ifx \showarticletitle \undefined \def \showarticletitle #1{#1}   \fi
\ifx \showURL      \undefined \def \showURL       {\relax}        \fi
% invisible to TeX
\providecommand\bibfield[2]{#2}
\providecommand\bibinfo[2]{#2}
\providecommand\natexlab[1]{#1}
\providecommand\showeprint[2][]{arXiv:#2}

\bibitem[\protect\citeauthoryear{Basilico and Hofmann}{Basilico and
  Hofmann}{2004}]%
        {Basilico:2004:UCC:1015330.1015394}
\bibfield{author}{\bibinfo{person}{Justin Basilico} {and}
  \bibinfo{person}{Thomas Hofmann}.} \bibinfo{year}{2004}\natexlab{}.
\newblock \showarticletitle{Unifying Collaborative and Content-based
  Filtering}. In \bibinfo{booktitle}{{\em Proceedings of the Twenty-first
  International Conference on Machine Learning}} {\em (\bibinfo{series}{ICML
  '04})}. \bibinfo{publisher}{ACM}, \bibinfo{address}{New York, NY, USA},
  \bibinfo{pages}{9--}.
\newblock
\showISBNx{1-58113-838-5}


\bibitem[\protect\citeauthoryear{Burke}{Burke}{2002}]%
        {Burke:2002:HRS:586321.586352}
\bibfield{author}{\bibinfo{person}{Robin Burke}.}
  \bibinfo{year}{2002}\natexlab{}.
\newblock \showarticletitle{Hybrid Recommender Systems: Survey and
  Experiments}.
\newblock \bibinfo{journal}{{\em User Modeling and User-Adapted Interaction\/}}
  \bibinfo{volume}{12}, \bibinfo{number}{4} (\bibinfo{date}{Nov.}
  \bibinfo{year}{2002}), \bibinfo{pages}{331--370}.
\newblock
\showISSN{0924-1868}


\bibitem[\protect\citeauthoryear{Cho, van Merrienboer, G{\"{u}}l{\c{c}}ehre,
  Bougares, Schwenk, and Bengio}{Cho et~al\mbox{.}}{2014}]%
        {ref:choEtAl2014}
\bibfield{author}{\bibinfo{person}{Kyunghyun Cho}, \bibinfo{person}{Bart van
  Merrienboer}, \bibinfo{person}{{\c{C}}aglar G{\"{u}}l{\c{c}}ehre},
  \bibinfo{person}{Fethi Bougares}, \bibinfo{person}{Holger Schwenk}, {and}
  \bibinfo{person}{Yoshua Bengio}.} \bibinfo{year}{2014}\natexlab{}.
\newblock \showarticletitle{Learning Phrase Representations using {RNN}
  Encoder-Decoder for Statistical Machine Translation}.
\newblock \bibinfo{journal}{{\em CoRR\/}}  \bibinfo{volume}{abs/1406.1078}
  (\bibinfo{year}{2014}).
\newblock


\bibitem[\protect\citeauthoryear{Chung, G{\"{u}}l{\c{c}}ehre, Cho, and
  Bengio}{Chung et~al\mbox{.}}{2014}]%
        {ref:gru}
\bibfield{author}{\bibinfo{person}{Junyoung Chung},
  \bibinfo{person}{{\c{C}}aglar G{\"{u}}l{\c{c}}ehre},
  \bibinfo{person}{KyungHyun Cho}, {and} \bibinfo{person}{Yoshua Bengio}.}
  \bibinfo{year}{2014}\natexlab{}.
\newblock \showarticletitle{Empirical Evaluation of Gated Recurrent Neural
  Networks on Sequence Modeling}.
\newblock \bibinfo{journal}{{\em CoRR\/}}  \bibinfo{volume}{abs/1412.3555}
  (\bibinfo{year}{2014}).
\newblock


\bibitem[\protect\citeauthoryear{Covington, Adams, and Sargin}{Covington
  et~al\mbox{.}}{2016a}]%
        {ref:covingtonEtAl2016}
\bibfield{author}{\bibinfo{person}{Paul Covington}, \bibinfo{person}{Jay
  Adams}, {and} \bibinfo{person}{Emre Sargin}.}
  \bibinfo{year}{2016}\natexlab{a}.
\newblock \showarticletitle{Deep Neural Networks for YouTube Recommendations}.
  In \bibinfo{booktitle}{{\em Proceedings of the 10th ACM Conference on
  Recommender Systems}} {\em (\bibinfo{series}{RecSys '16})}.
  \bibinfo{publisher}{ACM}, \bibinfo{address}{New York, NY, USA},
  \bibinfo{pages}{191--198}.
\newblock
\showISBNx{978-1-4503-4035-9}


\bibitem[\protect\citeauthoryear{Covington, Adams, and Sargin}{Covington
  et~al\mbox{.}}{2016b}]%
        {ref:youtube}
\bibfield{author}{\bibinfo{person}{Paul Covington}, \bibinfo{person}{Jay
  Adams}, {and} \bibinfo{person}{Emre Sargin}.}
  \bibinfo{year}{2016}\natexlab{b}.
\newblock \showarticletitle{Deep Neural Networks for YouTube Recommendations}.
  In \bibinfo{booktitle}{{\em Proceedings of the 10th ACM Conference on
  Recommender Systems}}. \bibinfo{address}{New York, NY, USA}.
\newblock


\bibitem[\protect\citeauthoryear{Deshpande and Karypis}{Deshpande and
  Karypis}{2004}]%
        {ref:deshpande2004}
\bibfield{author}{\bibinfo{person}{Mukund Deshpande} {and}
  \bibinfo{person}{George Karypis}.} \bibinfo{year}{2004}\natexlab{}.
\newblock \showarticletitle{Item-based top-N Recommendation Algorithms}.
\newblock \bibinfo{journal}{{\em ACM Trans. Inf. Syst.\/}}
  \bibinfo{volume}{22}, \bibinfo{number}{1} (\bibinfo{date}{Jan.}
  \bibinfo{year}{2004}), \bibinfo{pages}{143--177}.
\newblock
\showISSN{1046-8188}


\bibitem[\protect\citeauthoryear{Dieleman}{Dieleman}{2014}]%
        {ref:spotify}
\bibfield{author}{\bibinfo{person}{Sander Dieleman}.}
  \bibinfo{year}{2014}\natexlab{}.
\newblock \bibinfo{title}{Recommending Music on Spotify with Deep Learning}.
\newblock
  \bibinfo{howpublished}{\url{http://benanne.github.io/2014/08/05/spotify-cnns.html}}.
    (\bibinfo{year}{2014}).
\newblock


\bibitem[\protect\citeauthoryear{Hochreiter and Schmidhuber}{Hochreiter and
  Schmidhuber}{1997}]%
        {ref:lstm}
\bibfield{author}{\bibinfo{person}{Sepp Hochreiter} {and}
  \bibinfo{person}{J\"{u}rgen Schmidhuber}.} \bibinfo{year}{1997}\natexlab{}.
\newblock \showarticletitle{Long Short-Term Memory}.
\newblock \bibinfo{journal}{{\em Neural Comput.\/}} \bibinfo{volume}{9},
  \bibinfo{number}{8} (\bibinfo{date}{Nov.} \bibinfo{year}{1997}),
  \bibinfo{pages}{1735--1780}.
\newblock
\showISSN{0899-7667}


\bibitem[\protect\citeauthoryear{Hu, Lu, Li, and Chen}{Hu
  et~al\mbox{.}}{2014}]%
        {NIPS2014_5550}
\bibfield{author}{\bibinfo{person}{Baotian Hu}, \bibinfo{person}{Zhengdong Lu},
  \bibinfo{person}{Hang Li}, {and} \bibinfo{person}{Qingcai Chen}.}
  \bibinfo{year}{2014}\natexlab{}.
\newblock \showarticletitle{Convolutional Neural Network Architectures for
  Matching Natural Language Sentences}.
\newblock In \bibinfo{booktitle}{{\em Advances in Neural Information Processing
  Systems 27}}, \bibfield{editor}{\bibinfo{person}{Z.~Ghahramani},
  \bibinfo{person}{M.~Welling}, \bibinfo{person}{C.~Cortes},
  \bibinfo{person}{N.~D. Lawrence}, {and} \bibinfo{person}{K.~Q. Weinberger}}
  (Eds.). \bibinfo{publisher}{Curran Associates, Inc.},
  \bibinfo{pages}{2042--2050}.
\newblock


\bibitem[\protect\citeauthoryear{Hu, Koren, and Volinsky}{Hu
  et~al\mbox{.}}{2008}]%
        {ref:huEtAl2008}
\bibfield{author}{\bibinfo{person}{Yifan Hu}, \bibinfo{person}{Yehuda Koren},
  {and} \bibinfo{person}{Chris Volinsky}.} \bibinfo{year}{2008}\natexlab{}.
\newblock \showarticletitle{Collaborative Filtering for Implicit Feedback
  Datasets}. In \bibinfo{booktitle}{{\em Proceedings of the 2008 Eighth IEEE
  International Conference on Data Mining}} {\em (\bibinfo{series}{ICDM '08})}.
  \bibinfo{publisher}{IEEE Computer Society}, \bibinfo{address}{Washington, DC,
  USA}, \bibinfo{pages}{263--272}.
\newblock
\showISBNx{978-0-7695-3502-9}


\bibitem[\protect\citeauthoryear{Huang, He, Gao, Deng, Acero, and Heck}{Huang
  et~al\mbox{.}}{2013}]%
        {Huang:2013:LDS:2505515.2505665}
\bibfield{author}{\bibinfo{person}{Po-Sen Huang}, \bibinfo{person}{Xiaodong
  He}, \bibinfo{person}{Jianfeng Gao}, \bibinfo{person}{Li Deng},
  \bibinfo{person}{Alex Acero}, {and} \bibinfo{person}{Larry Heck}.}
  \bibinfo{year}{2013}\natexlab{}.
\newblock \showarticletitle{Learning Deep Structured Semantic Models for Web
  Search Using Clickthrough Data}. In \bibinfo{booktitle}{{\em Proceedings of
  the 22Nd ACM International Conference on Information \& Knowledge
  Management}} {\em (\bibinfo{series}{CIKM '13})}. \bibinfo{publisher}{ACM},
  \bibinfo{address}{New York, NY, USA}, \bibinfo{pages}{2333--2338}.
\newblock
\showISBNx{978-1-4503-2263-8}


\bibitem[\protect\citeauthoryear{Jaech, Kamisetty, Ringger, and Clarke}{Jaech
  et~al\mbox{.}}{2017}]%
        {DBLP:journals/corr/JaechKRC17}
\bibfield{author}{\bibinfo{person}{Aaron Jaech}, \bibinfo{person}{Hetunandan
  Kamisetty}, \bibinfo{person}{Eric~K. Ringger}, {and} \bibinfo{person}{Charlie
  Clarke}.} \bibinfo{year}{2017}\natexlab{}.
\newblock \showarticletitle{Match-Tensor: a Deep Relevance Model for Search}.
\newblock \bibinfo{journal}{{\em CoRR\/}}  \bibinfo{volume}{abs/1701.07795}
  (\bibinfo{year}{2017}).
\newblock


\bibitem[\protect\citeauthoryear{Jannach, Zanker, Felfernig, and
  Friedrich}{Jannach et~al\mbox{.}}{2010}]%
        {ref:jannachEtAl2010}
\bibfield{author}{\bibinfo{person}{Dietmar Jannach}, \bibinfo{person}{Markus
  Zanker}, \bibinfo{person}{Alexander Felfernig}, {and}
  \bibinfo{person}{Gerhard Friedrich}.} \bibinfo{year}{2010}\natexlab{}.
\newblock \bibinfo{booktitle}{{\em Recommender Systems: An Introduction\/}
  (\bibinfo{edition}{1st} ed.)}.
\newblock \bibinfo{publisher}{Cambridge University Press},
  \bibinfo{address}{New York, NY, USA}.
\newblock
\showISBNx{0521493366, 9780521493369}


\bibitem[\protect\citeauthoryear{Jing, Liu, Kislyuk, Zhai, Xu, Donahue, and
  Tavel}{Jing et~al\mbox{.}}{2015}]%
        {Jing:2015:VSP:2783258.2788621}
\bibfield{author}{\bibinfo{person}{Yushi Jing}, \bibinfo{person}{David Liu},
  \bibinfo{person}{Dmitry Kislyuk}, \bibinfo{person}{Andrew Zhai},
  \bibinfo{person}{Jiajing Xu}, \bibinfo{person}{Jeff Donahue}, {and}
  \bibinfo{person}{Sarah Tavel}.} \bibinfo{year}{2015}\natexlab{}.
\newblock \showarticletitle{Visual Search at Pinterest}. In
  \bibinfo{booktitle}{{\em Proceedings of the 21th ACM SIGKDD International
  Conference on Knowledge Discovery and Data Mining}} {\em
  (\bibinfo{series}{KDD '15})}. \bibinfo{publisher}{ACM}, \bibinfo{address}{New
  York, NY, USA}, \bibinfo{pages}{1889--1898}.
\newblock
\showISBNx{978-1-4503-3664-2}


\bibitem[\protect\citeauthoryear{Koren, Bell, and Volinsky}{Koren
  et~al\mbox{.}}{2009}]%
        {ref:korenEtAl2009}
\bibfield{author}{\bibinfo{person}{Yehuda Koren}, \bibinfo{person}{Robert
  Bell}, {and} \bibinfo{person}{Chris Volinsky}.}
  \bibinfo{year}{2009}\natexlab{}.
\newblock \showarticletitle{Matrix Factorization Techniques for Recommender
  Systems}.
\newblock \bibinfo{journal}{{\em Computer\/}} \bibinfo{volume}{42},
  \bibinfo{number}{8} (\bibinfo{date}{Aug.} \bibinfo{year}{2009}),
  \bibinfo{pages}{30--37}.
\newblock
\showISSN{0018-9162}


\bibitem[\protect\citeauthoryear{Li and Kim}{Li and Kim}{2003}]%
        {Li:2003:ACC:1118935.1118938}
\bibfield{author}{\bibinfo{person}{Qing Li} {and} \bibinfo{person}{Byeong~Man
  Kim}.} \bibinfo{year}{2003}\natexlab{}.
\newblock \showarticletitle{An Approach for Combining Content-based and
  Collaborative Filters}. In \bibinfo{booktitle}{{\em Proceedings of the Sixth
  International Workshop on Information Retrieval with Asian Languages - Volume
  11}} {\em (\bibinfo{series}{AsianIR '03})}. \bibinfo{publisher}{Association
  for Computational Linguistics}, \bibinfo{address}{Stroudsburg, PA, USA},
  \bibinfo{pages}{17--24}.
\newblock


\bibitem[\protect\citeauthoryear{Liang and Baldwin}{Liang and Baldwin}{2015}]%
        {Liang:2015:PRA:2806416.2806633}
\bibfield{author}{\bibinfo{person}{Huizhi Liang} {and} \bibinfo{person}{Timothy
  Baldwin}.} \bibinfo{year}{2015}\natexlab{}.
\newblock \showarticletitle{A Probabilistic Rating Auto-encoder for
  Personalized Recommender Systems}. In \bibinfo{booktitle}{{\em Proceedings of
  the 24th ACM International on Conference on Information and Knowledge
  Management}} {\em (\bibinfo{series}{CIKM '15})}. \bibinfo{publisher}{ACM},
  \bibinfo{address}{New York, NY, USA}, \bibinfo{pages}{1863--1866}.
\newblock
\showISBNx{978-1-4503-3794-6}


\bibitem[\protect\citeauthoryear{Linden, Smith, and York}{Linden
  et~al\mbox{.}}{2003}]%
        {ref:linden2003amazon}
\bibfield{author}{\bibinfo{person}{Greg Linden}, \bibinfo{person}{Brent Smith},
  {and} \bibinfo{person}{Jeremy York}.} \bibinfo{year}{2003}\natexlab{}.
\newblock \showarticletitle{Amazon. com recommendations: Item-to-item
  collaborative filtering}.
\newblock \bibinfo{journal}{{\em IEEE Internet computing\/}}
  \bibinfo{volume}{7}, \bibinfo{number}{1} (\bibinfo{year}{2003}),
  \bibinfo{pages}{76--80}.
\newblock


\bibitem[\protect\citeauthoryear{Lops, de~Gemmis, and Semeraro}{Lops
  et~al\mbox{.}}{2011}]%
        {reference:rsh:LopsGS11}
\bibfield{author}{\bibinfo{person}{Pasquale Lops}, \bibinfo{person}{Marco de
  Gemmis}, {and} \bibinfo{person}{Giovanni Semeraro}.}
  \bibinfo{year}{2011}\natexlab{}.
\newblock \showarticletitle{Content-based Recommender Systems: State of the Art
  and Trends.}
\newblock In \bibinfo{booktitle}{{\em Recommender Systems Handbook}},
  \bibfield{editor}{\bibinfo{person}{Francesco Ricci}, \bibinfo{person}{Lior
  Rokach}, \bibinfo{person}{Bracha Shapira}, {and} \bibinfo{person}{Paul~B.
  Kantor}} (Eds.). \bibinfo{publisher}{Springer}, \bibinfo{pages}{73--105}.
\newblock
\showISBNx{978-0-387-85819-7}


\bibitem[\protect\citeauthoryear{Lu and Li}{Lu and Li}{2013}]%
        {NIPS2013_5019}
\bibfield{author}{\bibinfo{person}{Zhengdong Lu} {and} \bibinfo{person}{Hang
  Li}.} \bibinfo{year}{2013}\natexlab{}.
\newblock \showarticletitle{A Deep Architecture for Matching Short Texts}.
\newblock In \bibinfo{booktitle}{{\em Advances in Neural Information Processing
  Systems 26}}, \bibfield{editor}{\bibinfo{person}{C.~J.~C. Burges},
  \bibinfo{person}{L.~Bottou}, \bibinfo{person}{M.~Welling},
  \bibinfo{person}{Z.~Ghahramani}, {and} \bibinfo{person}{K.~Q. Weinberger}}
  (Eds.). \bibinfo{publisher}{Curran Associates, Inc.},
  \bibinfo{pages}{1367--1375}.
\newblock


\bibitem[\protect\citeauthoryear{McAuley, Targett, Shi, and van~den
  Hengel}{McAuley et~al\mbox{.}}{2015}]%
        {McAuley:2015:IRS:2766462.2767755}
\bibfield{author}{\bibinfo{person}{Julian McAuley},
  \bibinfo{person}{Christopher Targett}, \bibinfo{person}{Qinfeng Shi}, {and}
  \bibinfo{person}{Anton van~den Hengel}.} \bibinfo{year}{2015}\natexlab{}.
\newblock \showarticletitle{Image-Based Recommendations on Styles and
  Substitutes}. In \bibinfo{booktitle}{{\em Proceedings of the 38th
  International ACM SIGIR Conference on Research and Development in Information
  Retrieval}} {\em (\bibinfo{series}{SIGIR '15})}. \bibinfo{publisher}{ACM},
  \bibinfo{address}{New York, NY, USA}, \bibinfo{pages}{43--52}.
\newblock
\showISBNx{978-1-4503-3621-5}


\bibitem[\protect\citeauthoryear{Melville, Mooney, and Nagarajan}{Melville
  et~al\mbox{.}}{2002}]%
        {Melville:2002:CCF:777092.777124}
\bibfield{author}{\bibinfo{person}{Prem Melville}, \bibinfo{person}{Raymod~J.
  Mooney}, {and} \bibinfo{person}{Ramadass Nagarajan}.}
  \bibinfo{year}{2002}\natexlab{}.
\newblock \showarticletitle{Content-boosted Collaborative Filtering for
  Improved Recommendations}. In \bibinfo{booktitle}{{\em Eighteenth National
  Conference on Artificial Intelligence}}. \bibinfo{publisher}{American
  Association for Artificial Intelligence}, \bibinfo{address}{Menlo Park, CA,
  USA}, \bibinfo{pages}{187--192}.
\newblock
\showISBNx{0-262-51129-0}


\bibitem[\protect\citeauthoryear{Mikolov, Chen, Corrado, and Dean}{Mikolov
  et~al\mbox{.}}{2013a}]%
        {ref:mikolovEtAl2013a}
\bibfield{author}{\bibinfo{person}{Tomas Mikolov}, \bibinfo{person}{Kai Chen},
  \bibinfo{person}{Greg Corrado}, {and} \bibinfo{person}{Jeffrey Dean}.}
  \bibinfo{year}{2013}\natexlab{a}.
\newblock \showarticletitle{Efficient Estimation of Word Representations in
  Vector Space}.
\newblock \bibinfo{journal}{{\em CoRR\/}}  \bibinfo{volume}{abs/1301.3781}
  (\bibinfo{year}{2013}).
\newblock


\bibitem[\protect\citeauthoryear{Mikolov, Sutskever, Chen, Corrado, and
  Dean}{Mikolov et~al\mbox{.}}{2013b}]%
        {ref:mikolovEtAl2013b}
\bibfield{author}{\bibinfo{person}{Tomas Mikolov}, \bibinfo{person}{Ilya
  Sutskever}, \bibinfo{person}{Kai Chen}, \bibinfo{person}{Greg Corrado}, {and}
  \bibinfo{person}{Jeffrey Dean}.} \bibinfo{year}{2013}\natexlab{b}.
\newblock \showarticletitle{Distributed Representations of Words and Phrases
  and Their Compositionality}. In \bibinfo{booktitle}{{\em Proceedings of the
  26th International Conference on Neural Information Processing Systems}} {\em
  (\bibinfo{series}{NIPS'13})}. \bibinfo{publisher}{Curran Associates Inc.},
  \bibinfo{address}{USA}, \bibinfo{pages}{3111--3119}.
\newblock


\bibitem[\protect\citeauthoryear{Mitra, Diaz, and Craswell}{Mitra
  et~al\mbox{.}}{2016}]%
        {DBLP:journals/corr/Mitra0C16}
\bibfield{author}{\bibinfo{person}{Bhaskar Mitra}, \bibinfo{person}{Fernando
  Diaz}, {and} \bibinfo{person}{Nick Craswell}.}
  \bibinfo{year}{2016}\natexlab{}.
\newblock \showarticletitle{Learning to Match Using Local and Distributed
  Representations of Text for Web Search}.
\newblock \bibinfo{journal}{{\em CoRR\/}}  \bibinfo{volume}{abs/1610.08136}
  (\bibinfo{year}{2016}).
\newblock


\bibitem[\protect\citeauthoryear{Popescul, Pennock, and Lawrence}{Popescul
  et~al\mbox{.}}{2001}]%
        {Popescul:2001:PMU:2074022.2074076}
\bibfield{author}{\bibinfo{person}{Alexandrin Popescul},
  \bibinfo{person}{David~M. Pennock}, {and} \bibinfo{person}{Steve Lawrence}.}
  \bibinfo{year}{2001}\natexlab{}.
\newblock \showarticletitle{Probabilistic Models for Unified Collaborative and
  Content-based Recommendation in Sparse-data Environments}. In
  \bibinfo{booktitle}{{\em Proceedings of the Seventeenth Conference on
  Uncertainty in Artificial Intelligence}} {\em (\bibinfo{series}{UAI'01})}.
  \bibinfo{publisher}{Morgan Kaufmann Publishers Inc.}, \bibinfo{address}{San
  Francisco, CA, USA}, \bibinfo{pages}{437--444}.
\newblock
\showISBNx{1-55860-800-1}


\bibitem[\protect\citeauthoryear{Salakhutdinov and Mnih}{Salakhutdinov and
  Mnih}{2008}]%
        {ref:salakhutdinovMinh2011}
\bibfield{author}{\bibinfo{person}{Ruslan Salakhutdinov} {and}
  \bibinfo{person}{Andriy Mnih}.} \bibinfo{year}{2008}\natexlab{}.
\newblock \showarticletitle{Probabilistic Matrix Factorization}. In
  \bibinfo{booktitle}{{\em Advances in Neural Information Processing Systems}},
  Vol.~\bibinfo{volume}{20}.
\newblock


\bibitem[\protect\citeauthoryear{Severyn and Moschitti}{Severyn and
  Moschitti}{2015}]%
        {Severyn:2015:LRS:2766462.2767738}
\bibfield{author}{\bibinfo{person}{Aliaksei Severyn} {and}
  \bibinfo{person}{Alessandro Moschitti}.} \bibinfo{year}{2015}\natexlab{}.
\newblock \showarticletitle{Learning to Rank Short Text Pairs with
  Convolutional Deep Neural Networks}. In \bibinfo{booktitle}{{\em Proceedings
  of the 38th International ACM SIGIR Conference on Research and Development in
  Information Retrieval}} {\em (\bibinfo{series}{SIGIR '15})}.
  \bibinfo{publisher}{ACM}, \bibinfo{address}{New York, NY, USA},
  \bibinfo{pages}{373--382}.
\newblock
\showISBNx{978-1-4503-3621-5}


\bibitem[\protect\citeauthoryear{Shen, He, Gao, Deng, and Mesnil}{Shen
  et~al\mbox{.}}{2014}]%
        {Shen:2014:LSM:2661829.2661935}
\bibfield{author}{\bibinfo{person}{Yelong Shen}, \bibinfo{person}{Xiaodong He},
  \bibinfo{person}{Jianfeng Gao}, \bibinfo{person}{Li Deng}, {and}
  \bibinfo{person}{Gr{\'e}goire Mesnil}.} \bibinfo{year}{2014}\natexlab{}.
\newblock \showarticletitle{A Latent Semantic Model with Convolutional-Pooling
  Structure for Information Retrieval}. In \bibinfo{booktitle}{{\em Proceedings
  of the 23rd ACM International Conference on Conference on Information and
  Knowledge Management}} {\em (\bibinfo{series}{CIKM '14})}.
  \bibinfo{publisher}{ACM}, \bibinfo{address}{New York, NY, USA},
  \bibinfo{pages}{101--110}.
\newblock
\showISBNx{978-1-4503-2598-1}


\bibitem[\protect\citeauthoryear{{Theano Development Team}}{{Theano Development
  Team}}{2016}]%
        {ref:theano}
\bibfield{author}{\bibinfo{person}{{Theano Development Team}}.}
  \bibinfo{year}{2016}\natexlab{}.
\newblock \showarticletitle{{Theano: A {Python} framework for fast computation
  of mathematical expressions}}.
\newblock \bibinfo{journal}{{\em arXiv e-prints\/}}
  \bibinfo{volume}{abs/1605.02688} (\bibinfo{date}{May} \bibinfo{year}{2016}).
\newblock


\bibitem[\protect\citeauthoryear{Xu, Chen, Lukasiewicz, Miao, and Meng}{Xu
  et~al\mbox{.}}{2016}]%
        {Xu:2016:TPR:2983323.2983874}
\bibfield{author}{\bibinfo{person}{Zhenghua Xu}, \bibinfo{person}{Cheng Chen},
  \bibinfo{person}{Thomas Lukasiewicz}, \bibinfo{person}{Yishu Miao}, {and}
  \bibinfo{person}{Xiangwu Meng}.} \bibinfo{year}{2016}\natexlab{}.
\newblock \showarticletitle{Tag-Aware Personalized Recommendation Using a
  Deep-Semantic Similarity Model with Negative Sampling}. In
  \bibinfo{booktitle}{{\em Proceedings of the 25th ACM International on
  Conference on Information and Knowledge Management}} {\em
  (\bibinfo{series}{CIKM '16})}. \bibinfo{publisher}{ACM},
  \bibinfo{address}{New York, NY, USA}, \bibinfo{pages}{1921--1924}.
\newblock
\showISBNx{978-1-4503-4073-1}


\bibitem[\protect\citeauthoryear{Zhang, Gong, Wu, Huang, and Huang}{Zhang
  et~al\mbox{.}}{2016}]%
        {Zhang:2016:RPA:2983323.2983809}
\bibfield{author}{\bibinfo{person}{Qi Zhang}, \bibinfo{person}{Yeyun Gong},
  \bibinfo{person}{Jindou Wu}, \bibinfo{person}{Haoran Huang}, {and}
  \bibinfo{person}{Xuanjing Huang}.} \bibinfo{year}{2016}\natexlab{}.
\newblock \showarticletitle{Retweet Prediction with Attention-based Deep Neural
  Network}. In \bibinfo{booktitle}{{\em Proceedings of the 25th ACM
  International on Conference on Information and Knowledge Management}} {\em
  (\bibinfo{series}{CIKM '16})}. \bibinfo{publisher}{ACM},
  \bibinfo{address}{New York, NY, USA}, \bibinfo{pages}{75--84}.
\newblock
\showISBNx{978-1-4503-4073-1}


\bibitem[\protect\citeauthoryear{Zhang, Zhao, and LeCun}{Zhang
  et~al\mbox{.}}{2015}]%
        {Zhang:2015:CCN:2969239.2969312}
\bibfield{author}{\bibinfo{person}{Xiang Zhang}, \bibinfo{person}{Junbo Zhao},
  {and} \bibinfo{person}{Yann LeCun}.} \bibinfo{year}{2015}\natexlab{}.
\newblock \showarticletitle{Character-level Convolutional Networks for Text
  Classification}. In \bibinfo{booktitle}{{\em Proceedings of the 28th
  International Conference on Neural Information Processing Systems}} {\em
  (\bibinfo{series}{NIPS'15})}. \bibinfo{publisher}{MIT Press},
  \bibinfo{address}{Cambridge, MA, USA}, \bibinfo{pages}{649--657}.
\newblock


\bibitem[\protect\citeauthoryear{Zuo, Zeng, Gong, and Jiao}{Zuo
  et~al\mbox{.}}{2016}]%
        {Zuo:2016:TRS:2955826.2955877}
\bibfield{author}{\bibinfo{person}{Yi Zuo}, \bibinfo{person}{Jiulin Zeng},
  \bibinfo{person}{Maoguo Gong}, {and} \bibinfo{person}{Licheng Jiao}.}
  \bibinfo{year}{2016}\natexlab{}.
\newblock \showarticletitle{Tag-aware Recommender Systems Based on Deep Neural
  Networks}.
\newblock \bibinfo{journal}{{\em Neurocomput.\/}} \bibinfo{volume}{204},
  \bibinfo{number}{C} (\bibinfo{date}{Sept.} \bibinfo{year}{2016}),
  \bibinfo{pages}{51--60}.
\newblock
\showISSN{0925-2312}


\end{thebibliography}

\end{document}